\documentclass[conference]{IEEEtran}
\usepackage{lipsum}
\usepackage{flushend}
\usepackage{ifpdf}
\usepackage{verbatim}
\usepackage{cite}
\usepackage[pdftex]{graphicx}
\usepackage[cmex10]{amsmath}
\usepackage[plain]{algorithm}
\usepackage{algpseudocode}
\makeatletter
\def\BState{\State\hskip-\ALG@thistlm}
\makeatother
\usepackage{algorithmicx}
\usepackage{setspace}
\usepackage{array}
\usepackage{mdwmath}
\usepackage{mdwtab}
\usepackage{eqparbox}
\usepackage{url}
\usepackage{xcolor}
\usepackage{epstopdf}
\usepackage{paralist}
\usepackage{caption}
\usepackage{listings}
\usepackage{color}
\usepackage{xr}
\usepackage{amssymb}
\usepackage{textcomp}
\usepackage{multirow}
\usepackage[mathscr]{euscript}
\usepackage{hhline}
\usepackage{fancyvrb}
\usepackage{relsize}
\usepackage{lipsum}
\usepackage{array}
\usepackage{csquotes}
\usepackage{listings}
\lstdefinelanguage
   [x64]{Assembler}     % add a "x64" dialect of Assembler
   [x86masm]{Assembler} % based on the "x86masm" dialect
   {morekeywords={CDQE,CQO,CMPSQ,CMPXCHG16B,JRCXZ,LODSQ,MOVSXD, %
                  POPFQ,PUSHFQ,SCASQ,STOSQ,IRETQ,RDTSCP,SWAPGS, %
                  TEST,JLE,JE,AND,XOR,CMP,SUB,NOPL,XORPS,MOVSD,
                  MOVL,CVTSI2SD,LEA,XCHG,INC,DEC,ADD,RET,RETQ,
                  CLTQ,
                  eax,edx,ecx,ebx,esi,edi,esp,ebp,
                  rax,rdx,rcx,rbx,rsi,rdi,rsp,rbp,
                  define,store,gep,trunc,icmp,phi,sitofp,br,
                  xmm0,xmm1,void,for,double,unsigned}} % etc.

\author{\IEEEauthorblockN{Vishal Chandra Sharma, Ganesh Gopalakrishnan}
\IEEEauthorblockA{School of Computing, University of Utah\\
Email: \{vcsharma,ganesh\}@cs.utah.edu}
\and
\IEEEauthorblockN{Sriram Krishnamoorthy}
\IEEEauthorblockA{Pacific Northwest National Laboratory \\
Email: sriram@pnnl.gov}
}

\begin{document}
\title{PRESAGE: Protecting Structured Address Generation against Soft Errors}
\maketitle
\begin{abstract}
Modern computer scaling trends in pursuit of larger component 
counts and power efficiency have, unfortunately, lead to less
reliable hardware and consequently soft errors escaping into
application data (``silent data corruptions'').
Techniques to enhance system resilience hinge on the availability
of efficient error detectors that have high detection rates,
low false positive rates, and lower computational overhead.
Unfortunately, efficient detectors to detect faults during
address generation (to index large arrays) have not been 
widely researched.
We present a novel lightweight compiler-driven technique called
PRESAGE for detecting bit-flips affecting structured address computations.
A key insight underlying PRESAGE is that any address computation
scheme that flows an already incurred error is better than a scheme
that corrupts one particular array access but otherwise (falsely) appears
to compute perfectly.
Enabling the flow of errors allows one to situate detectors at loop exit points,
and helps turn silent corruptions into easily detectable error situations.
Our experiments using PolyBench benchmark suite indicate that
PRESAGE-based error detectors have a high error-detection rate
while incurring low overheads.

\end{abstract}

\section{\textbf{Introduction}}
\label{sec:intro}

High performance computing (HPC) applications will soon be running at very 
high scales on systems with large component counts.
The shrinking dimensions and reducing power requirements of the 
transistors used in the memory elements of these massively parallel 
systems make them increasingly vulnerable to temporary bit-flips 
induced by energetic particle strikes such as alpha particle and 
cosmic rays.
These temporary bit-flips occurring in memory elements are 
often referred as soft errors.
Previous studies project an upward trend in soft error induced 
vulnerabilities in HPC systems thereby pushing down
their Mean-Time-To-Failure (MTTF)\cite{megastudy09,megastudy14}.

These trends drastically increase the likelihood of a 
bit-flip occurring in long-lived computations.
Specifically, a bit-flip affecting computational states of a program under 
execution such as ALU operations or live register values, 
may lead to a silent data corruption (SDC) in the 
final program output.
Making the matter worse, such erroneous values may propagate 
to multiple compute nodes in massively parallel HPC 
systems~\cite{sc15bose}.

The key focus of this paper is to detect bit-flips
affecting address computation of array elements.
For example, to load a value stored in an array $A$ 
at an index \texttt{i}, a compiler must first compute the 
address of the location referred by the index \texttt{i}.
A compiler performs this operation under-the-hood
by using the base address of $A$ and adding to it
an offset value computed using the index \texttt{i}.
This style of address generation scheme which uses a 
base address and an offset to generate the destination
address is often referred as the \emph{structured address
generation}.
Accordingly, the computations done in the context of a
\emph{structured address generation} are referred
as \emph{structured address computations}.

Computational kernels used in HPC applications often 
involve array accesses inside loops thus requiring
\emph{structured address computations}.
For these kernels, there is a real chance of one of
their \emph{structured address computations} getting 
affected by a bit-flip.
A \emph{structured address computation}, pertaining
to an array, when subjected to a bit-flip may produce an 
incorrect address which still refers to a valid address in 
the address space of the array.
Using the value stored at this incorrect but a \emph{valid} 
address may lead to SDC without causing a program crash
or any other user-detectable-errors.

In this paper, we demonstrate that the bit-flips affecting 
\emph{structured address computations} for aforementioned 
class of computational kernels lead to non-trivial SDC rates.
We also present a novel technique for detecting bit-flips
impacting \emph{structured address computations}.
Given that the \emph{structured address computations} involve
arithmetic operations thus requiring the use of a CPU's computational
resources, we consider an error model where bit-flips affect 
ALU operations and CPU register files.
We assume DRAM and cache memory to be error-free which is a reasonable
assumption as they are often protected using ECC mechanism\cite{ecc84,ecc05,KimSGE15,KimSE15}.
We further limit the scope of our error model by considering only
those ALU operations and register values which correspond to
\emph{structured address computations}.

Specifically, we make following contributions in
this paper:
\begin{enumerate}
 \item A fault injection driven study done on 10 benchmarks drawn from 
 PolyBench/C benchmark\cite{polywebref} demonstrating that the 
 \emph{structured address computations} in those benchmarks when
 subjected to bit-flips lead to non-trivial SDC rates.

 \item We present a novel scheme which employs instruction-level rewriting of 
 the address computation logic used in \emph{structured address computations}.
 This rewrite {\em preserves} an error in a \emph{structured address computation} 
 by intentionally corrupting all \emph{structured address computations} that follow it. 
 This requires creation of a dependency-chain between all \emph{structured address computation} 
 pertaining to a given array. Enabling the flow of error helps in following ways: 
 \begin{itemize}
  \item \textbf{Strategic Placement of Error Detectors}: Instead of checking each and every 
  \emph{structured address computations} for soft errors (which is prohibitively expensive), we 
  strategically place our error detectors at the end of a dependency chain. 
  \item \textbf{Promoting SDCs to Program Crashes}: By enabling the flow of error in 
  address computation logic, we increase the chances of promoting an SDC to a 
  \emph{user-visible} program crash.
 \end{itemize}
 \item We present a methodology for implementing our proposed scheme
 as a compiler-level technique called PRESAGE (\textbf{PR}ot\textbf{E}cting 
 \textbf{S}tructured \textbf{A}ddress \textbf{GE}neration).
 Specifically, we have implemented PRESAGE using LLVM compiler infrastructure 
 \cite{lattner2004,llvmwebref} as a transformation pass.
 LLVM preserves the pointer related information at LLVM intermediate representation
 (IR) level (as also highlighted in recent works \cite{DBLP:conf/dsn/WeiTLP14,everything-ptrs-santosh}) while providing the
 access to a rich set of application programming interfaces (APIs) for seamlessly 
 implementing PRESAGE transformations. 
 This is the key reason behind choosing LLVM as tool-of-choice.
\end{enumerate}
In summary, our error-detection approach is based on the following principle:

\begin{displayquote}
{\em The larger the fraction of system state an error corrupts, 
the easier it is to detect them.}
\end{displayquote} 
The rest of the paper is organized as follows. Sec.~\S\ref{sec:relwork}
provides a literary review of the closely related work done in this area. 
Sec.~\S\ref{sec:ex} explains the key idea through a set of small examples. 
Sec.~\S\ref{sec:method} formally introduces the key concepts and the 
methodology used for implementing PRESAGE. 
In Sec.~\S\ref{sec:result}, we provide a detailed analysis of the 
experimental results carried out to measure the efficacy of PRESAGE. 
Finally, Sec.~\S\ref{sec:con} summarizes they key takeaways and future 
directions for this work.

%---
\begin{figure*}[!htb]
\begin{minipage}{0.5\textwidth}
\begin{minipage}{1.0\textwidth}
\begin{lstlisting}[basicstyle=\footnotesize]
L0:  cmp    0x2,%esi
L1:  jl     L12
L2:  xor    %eax,%eax
L3:  mov    0x1,%ecx
L4:  xorps  %xmm0,%xmm0
L5:  cvtsi2sd %ecx,%xmm0
L6:  cltq   
L7:  movsd  %xmm0,(%rdi,%rax,8)
L8:  add    0x2,%eax
L9:  inc    %ecx
L10: cmp    %ecx,%esi
L11: jne    L4
L12: retq   
\end{lstlisting}
\vspace{-0.5em}
\caption{x86 representation of the foo() function}
\label{fig:foox86}
\end{minipage}
\end{minipage}
\begin{minipage}{0.5\textwidth}
\begin{minipage}{1.0\textwidth}
\begin{lstlisting}[basicstyle=\footnotesize]
L0: void foo1(double* a, unsigned n){
L1:   double* addr=a;
L2:   for(int i=1;i<n;i++){
L3:     int id=2*i-2;
L4:	addr=&a[id];
L5: 	*addr=i; 	
      }	
    }
\end{lstlisting}
\vspace{-1.em}
\caption{foo1() function}
\label{fig:foo1}
\end{minipage}
\begin{minipage}{1.0\textwidth}
\begin{lstlisting}[basicstyle=\footnotesize]
L0: void foo2(double *a, int n){   
L1:   double* addr=a; 
L2:   int pid=0;
L3:   for(int i=1;i<n;i++){
L4: 	int id=2*i-2;
L5: 	int rid=id-pid; 
L6: 	addr[rid]=i;
L7: 	pid=id;
L8: 	addr=&addr[rid];
      }	
    }
\end{lstlisting}
\vspace{-1.em}
\caption{foo2() function}
\label{fig:foo2}
\end{minipage}
\end{minipage}
\end{figure*}
%--end

%---

\section{\textbf{Background \& Related Work}}
\label{sec:relatedwork}
\label{sec:relwork}
A previous work by Casas-Guix et al.~\cite{casas2012FauResAlgMulSol} 
shows that an Algebraic Multigrid (AMG) solver is relatively immune 
to faults as they can often recover to an acceptable final answer even 
after encountering a momentary bit-flip in the data state.
They however realize that any fault in the space of pointers often
wreaks havoc, since the corrupted pointers tend to write data values
into intended memory spaces.
As a solution, they propose the use of pointer triplication, which not
only helps detect errors in the value of a pointer variable but also
correct the same.
Unfortunately, pointer triplication comes with a high overhead of runtime 
checks.
Also, they do not focus on the scenarios where corruptions in 
\emph{structured address computations} lead to SDC which is the 
key focus of our work.

Another work by Wei et al.\cite{DBLP:conf/dsn/WeiTLP14} highlights the 
difference between 
the results of the fault injection experiments done using a 
higher-level fault injector LLFI targeting instructions at LLVM IR level, 
and a lower-level, PIN based, fault injector performing fault injections
at x86 level.
This work highlights that LLVM offers a separate instruction called
\texttt{getElementPtr} for carrying out \emph{structured address 
computations} whereas at x86-level same instruction
can be used for computing address as well as performing 
non-address arithmetic computations.
Another recent work by Nagarakatte et al.~\cite{everything-ptrs-santosh} 
shows how by associating meta-data and by using Intel's recently introduced
MPX instructions, one can guard C/C++ programs against pointer-related
memory attacks.
The key portion of this work is also implemented using LLVM infrastructure.
The above two works, in a way, influenced our decision to choose 
LLVM for implementing PRESAGE.

Researchers have also explored the development of application-level error detectors 
for detecting soft-error affecting a program's control states\cite{Oh2002,kudialctes13,kulfi13}.
Another key area in application-level resilience is the algorithm based fault tolerance 
(ABFT) which exploits the algorithmic properties of well-known applications to derive
efficient error detectors\cite{sloan13,newsmhpdc16}.
Researchers have also focused in the past to optimize the placement of application
level error detectors at strategic program points.%
The information about these strategic location are usually derived through well established
static and dynamic program analysis techniques\cite{karthikprdc,shoestrg,dekruijf12,hari13}.
To the best of our knowledge, none of the previous works have focused on protecting 
\emph{structured address generation} leading to SDC which is the key focus of our work.

\section{\textbf{Motivating Example}}
\label{sec:example}
\label{sec:ex}
Fig.~\ref{fig:foo1} presents a simple C function \texttt{foo1} performing
store operations to even-indexed memory locations of an array \texttt{a[]} of size 
\texttt{2n} inside a \texttt{for} loop.
It also stores the last accessed array address into a variable \texttt{addr} at the
end of every loop-iteration.
Fig.~\ref{fig:foox86} represents the corresponding x86 code emitted 
for the \texttt{foo1} function  when compiled using clang compiler with 
\texttt{O1} optimization level.
Registers \texttt{\%esi} and \texttt{\%ecx} represent
the variable \texttt{n} and the loop iterator \texttt{i} of the function \texttt{foo1}
whereas registers \texttt{\%rdi} and \texttt{\%rax} correspond to the array's 
base address and index respectively. 
In every loop iteration, a destination array address is computed by the 
expression \texttt{(\%rdi,\%rax,0x8)} which evaluates to \texttt{(0x8*\%rax +\%rdi)}, 
the value in register \texttt{\%rax} is incremented by 2, and the base 
address stored in \texttt{\%rdi} remains fixed.
It is worth noting that the final address computation denoted by the expression 
\texttt{(\%rdi,\%rax,0x8)} is \emph{not user-visible} and is something compiler
does under-the-hood.

\begin{figure*}[!t]
\begin{minipage}{0.5\textwidth}
\centering
\includegraphics[width=6.5cm,height=4.5cm]{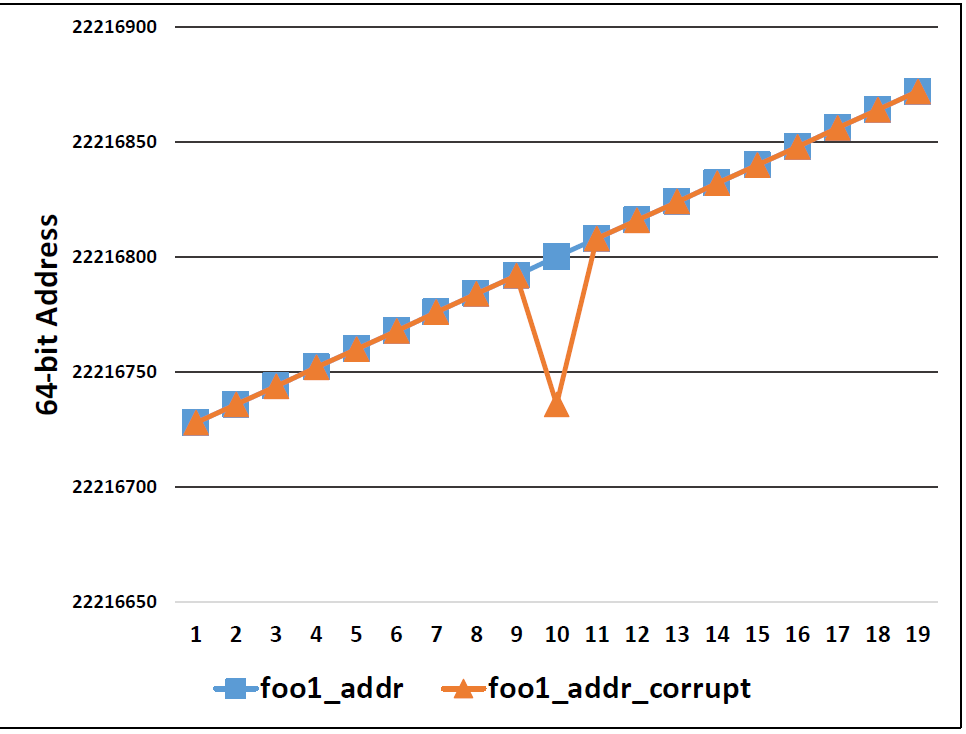}
\caption{Function \texttt{foo1} with no dependency-chains}
\label{fig:mex1}
\end{minipage}
\begin{minipage}{0.5\textwidth}
\centering
\includegraphics[width=6.5cm,height=4.5cm]{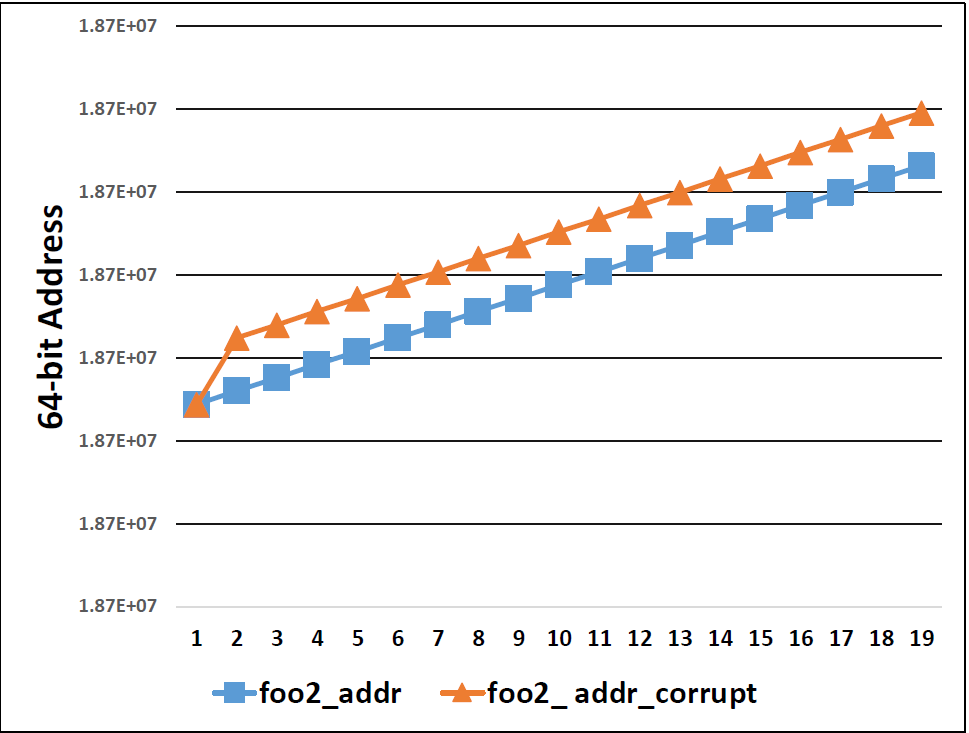}
\caption{Function \texttt{foo2} with a dependency-chain introduced in \texttt{addr}}
\label{fig:mex2}
\end{minipage}
\vspace{-2em}
\end{figure*}
In contrast to the fixed base address (FBA) scheme used in the function \texttt{foo1}, 
function \texttt{foo2} (shown in Fig.~\ref{fig:foo2}), a semantically 
equivalent version of \texttt{foo1}, introduces a novel relative base address (RBA) 
scheme.
Specifically, \texttt{foo2} uses an array address computed in a
loop iteration (\texttt{addr}) as the new base address for the 
next loop iteration along with a relative index (\texttt{rid}) as shown 
in Fig.~\ref{fig:foo1}.

This simple but powerful scheme creates a dependency chain in the address 
computation logic as the computation of any new address would depend on the 
last computed address.
Therefore, our RBA scheme guarantees that if an address 
computation of an array element gets corrupted then all subsequent address 
computations would also become erroneous.
This in turn enables us to strategically place error detectors at a handful
of places in a program (preferably at all program exit points) thereby
making the whole error detection process very lightweight.

For example, in functions \texttt{foo1} and \texttt{foo2}, 
the address of a new array element, computed during
every loop iteration, is stored in the variable \texttt{addr}.
The value stored in the variable \texttt{addr} may get corrupted
in following scenarios:
\noindent
\textbf{Error Scenario I}: A bit-flip occurs in the value stored 
in the loop-iterator variable \texttt{i} in functions \texttt{foo1} 
and \texttt{foo2}.\\
\noindent
\textbf{Error Scenario II}: A bit-flip affecting the value stored in 
the absolute index variable \texttt{id} in functions \texttt{foo1} 
and \texttt{foo2}.\\
\noindent
\textbf{Error Scenario III}: A bit-flip occurs in the value stored
in the relative index variable \texttt{rid} which is only present in 
the functions \texttt{foo2}.\\
\noindent
\textbf{Error Scenario IV}: A bit-flip affecting the value stored
in the variable \texttt{addr} in functions \texttt{foo1} and \texttt{foo2}.

The above program-level sites are listed in Table~\ref{tab:fs12}
for easy reference.
With respect to the error scenario IV, it is evident that 
only in the case of \texttt{foo2}, when the result of 
final address computation stored in \texttt{addr} is corrupted
during one of the loop iterations, all subsequent address computations 
in the remaining loop iterations would also get corrupted due to the 
dependency chain introduced in the address computation logic.
We further demonstrate the evidence of these dependency chains,
introduced by our RBA scheme, through a small set of fault injection 
driven experiments. 
Fig.~\ref{fig:mex1} presents the result of two independent runs
for function \texttt{foo1}.
The X-axis shows the number of loop iteration whereas the Y-axis
shows the value stored in the variable \texttt{addr}.
The execution with label \texttt{foo1\_addr} represents a fault-free
execution of \texttt{foo1}.
The execution with label \texttt{foo1\_addr\_corrupt} represents a faulty
execution of \texttt{foo1} where a single bit fault is introduced
at bit position $6$ of the value stored in \texttt{addr} during 
the tenth loop iteration.
Similarly, Fig.~\ref{fig:mex2} presents the result of two independent 
runs for function \texttt{foo2} such that a single bit fault is introduced
at bit position $6$ of the value stored in \texttt{addr} during 
the first loop iteration in the faulty execution represented by the label
\texttt{foo2\_addr\_corrupt}
We can clearly notice that only in the case of function \texttt{foo2},
once an address value stored in \texttt{addr}
gets corrupted, all subsequent address values stored in \texttt{addr} are
also corrupted.
\begin{table}[!h]
\begin{tabular}{| c | l | }
  \hline
  \textbf{Fault Site} & \textbf{Description}  \\
  \hline
  \texttt{i} & Loop iterator variable.  \\
  \hline
  \texttt{id} & Absolute index variable.  \\
  \hline
  \texttt{addr} & \shortstack[l]{A variable containing an address of a \\location in the array \texttt{a[]}.} \\
  \hline
  \texttt{rid} & \shortstack[l]{Relative index variable (only present in\\ \texttt{foo2}).} \\
  \hline
 \end{tabular}
\caption{List of fault sites in functions \texttt{foo1} and \texttt{foo2}}
\label{tab:fs12} 
\end{table}

\section{\textbf{Methodology}}
\label{sec:method}
\label{sec:method}
Sec. \S\ref{sec:ex} demonstrates that a simple rewrite of the address computation logic
introduces a dependency chain thereby enabling the flow of error.
Given that the address computation is often done in a user-transparent manner by a compiler, 
we implement our technique at the compiler-level.
Specifically, we choose LLVM compiler infrastructure to implement our 
technique as a transformation pass (hereon referred as PRESAGE) which works on LLVM's 
intermediate representation (IR).
Our implementation eliminates the need for any manual effort from programmers thereby 
allowing our technique to scale to non-trivial programs.
LLVM's intermediate representation (IR) provides a special instruction called 
\texttt{getelementptr} (hereon referred as \texttt{GEP} for brevity) for performing 
address computation of \texttt{Aggregate} 
types including \texttt{Array} type\footnote{LLVM's type system is explained in its 
language reference manual located at: \url{http://llvm.org/docs/LangRef.html}}.
Therefore, all analyses implemented as part of PRESAGE are centered
around the \texttt{GEP} instruction.
A \texttt{GEP} instruction requires a base address, one or more index 
values, and size of an element to compute an address and the computed address is 
often referred as \emph{structured address}. 
Given the key focus of our work is to protect these \emph{structured addresses}, 
the definition of an array on which PRESAGE transformations are applied closely follows 
the LLVM's \texttt{Array} type definition with some restrictions as explained below:

\noindent
\textbf{\emph{Definition 1}}: An \emph{array} in this paper always refers to a contiguous
arrangement of elements of the same \emph{type} laid out linearly in the memory. \\
\noindent 
\textbf{\emph{Definition 2}}: All structured address computations protected using PRESAGE must  
always use only one index for address computations. 
It is important to note that this is needed only to simplify the implementation and 
does not limit the scope of PRESAGE as multi-dimensional arrays can be easily represented 
using single-indexed scheme. 
For example, a two-dimensional array could be laid out linearly in memory by traversing it
in row-major or column-major fashion. \\
\noindent 
\textbf{\emph{Definition 3}}: The base addresses used in all structured address computations
protected by PRESAGE must be immutable. For example, if a PRESAGE transformation is applied on 
a callee function to protect its structured address computations then the callee function must 
not mutate those base addresses which are referenced in structured address computations 
protected by PRESAGE.\\
\noindent 
\textbf{\emph{Definition 4}}: Let $A$ be an array of arbitrary length and $A_i$ represents 
the $i$\emph{th} element of the array from its first element which starts with an index $0$.
The address $\gamma({A_i})$ of $A_i$ is computed using FBA scheme as shown in Eq.~\ref{eq:fba} 
where $\beta(A_i)$ represents the base address used to calculate the address of $A_i$, and $s_A$ 
denotes the size (in bytes) of the elements of the array $A$.
\begin{equation}
\gamma({A_i}) = \beta(A_i) + (s_A * i)
\label{eq:fba}
\end{equation}\\
\noindent 
\textbf{\emph{Definition 5}}: The address of $A_i$ when computed using our RBA scheme
uses a previously computed address $\gamma(A_j)$ as the new base address 
(hereafter referred as \emph{relative base}) and is denoted as $\gamma({A_{i\Box j}})$
as shown in Eq.~\ref{eq:rba}. The \emph{relative index} value used in Eq.~\ref{eq:rba}
is computed by simply subtracting the index value of the \emph{relative base} ($A_j$) from the 
index value of $A_i$.
In scenarios where the \emph{relative base} information is not known, the RBA scheme 
falls back to the FBA scheme for address computation.
\begin{equation}
\gamma({A_{i\Box j}}) = \gamma(A_j) + (s_A * (i-j))
\label{eq:rba}
\end{equation}

\noindent 
\textbf{\emph{Theorem}}: If we consider $A_i$ and $A_j$ as valid elements of an array
$A$, then $\gamma({A_{i\Box j}})$ $\equiv$ $\gamma({A_i})$ \emph{iff} 
$\beta({A_i}) \equiv \beta({A_j})$.

\noindent 
\textbf{\emph{Proof}}: Rewriting $\gamma(A_j)$ in Eq.~\ref{eq:rba} using Eq.~\ref{eq:fba}, 
we get Eq.~\ref{eq:eqv1}. 
\begin{equation}
\gamma({A_{i\Box j}}) = \beta(A_j) + (s_A * j) + (s_A * (i-j))
\label{eq:eqv1}
\end{equation}
By further simplying Eq.~\ref{eq:eqv1}, we finally get Eq.~\ref{eq:eqv2}. 
\begin{equation}
\gamma({A_{i\Box j}}) = \beta(A_j) + (s_A * i)
\label{eq:eqv2}
\end{equation}
Using Eqs.~\ref{eq:eqv2} and~\ref{eq:fba}, we get: $\gamma({A_{i\Box j}})$ 
$\equiv$ $\gamma({A_i})$ \emph{iff} $\beta({A_i}) \equiv \beta({A_j})$.

\begin{table}[!h]
\begin{tabular}{| l | l | }
  \hline
  \textbf{Term} & \textbf{Description}  \\
  \hline
  $\mathcal{F}$ & \shortstack[l]{A target function on which PRESAGE 
  \\transformations are applied.}  \\
  \hline
  $\mathbf{b}$ & \shortstack[l]{A base address with at least one user 
  \\in the target function.}  \\
  \hline
  $\mathcal{B}$ & \shortstack{A basic block in the target function.}  \\
  \hline
  $E(\mathcal{B}_1,\mathcal{B}_2)$ & \shortstack[l]{A boolean 
  function which returns \emph{true}\\ only if an edge exists from $\mathcal{B}_1$ 
  to $\mathcal{B}_2$.} \\	
  \hline  
  $\mathcal{L}_{\mathcal{B}_p}(\mathcal{B})$ & \shortstack[l]{A set of all immediate 
  predecessor basic\\ blocks of $\mathcal{B}$.}  \\
  \hline
  $\mathcal{L}_{\mathcal{B}_s}(\mathcal{B})$ & \shortstack[l]{A set of all immediate 
  successor basic\\ blocks of $\mathcal{B}$.}  \\
  \hline
  $\mathcal{L}_{\mathcal{B}_e}(\mathcal{F})$ & \shortstack[l]{A set of all exit  
  basic blocks\\ in the target function $\mathcal{F}$.}  \\
  \hline    
  $\mathcal{L}_\mathbf{b}(\mathcal{F})$ & \shortstack[l]{A set of all immutable base
  addresses\\in $\mathcal{F}$.}  \\
  \hline  
  $\mathcal{L}_{\mathcal{G}}(\mathcal{B},\mathbf{b} )$ & \shortstack[l]{A set of all 
  \texttt{GEP} instructions in $\mathcal{B}$ which\\use the base address $\mathbf{b}$.}\\
  \hline
  $\mathcal{M}_\phi$ & \shortstack[l]{A two-level nested hashmap with first key\\
  a basic block, second key a base address\\mapped to a \texttt{phi} node.}  \\
  \hline
  $\mathcal{M}_\mathcal{G}$ & \shortstack[l]{A two-level nested hashmap with first key\\
  a basic block, second key a base address\\mapped to a \texttt{GEP} instruction.}  \\
  \hline  
 \end{tabular}
\caption{Glossary of terms referred in this paper.}
\label{tab:terms} 
\end{table}

\subsection{\textbf{Error Model}}
We consider an error model where soft errors induce
a single-bit fault affecting CPU register files and 
ALU operations. 
We assume that memory elements such as data-cache and
DRAM are error free as they are usually protected using
ECC mechanism.
We implement our error model by targeting runtime instances 
of LLVM IR level instructions of a target function for 
fault injection.
For example, if there are $N$ dynamic IR-level instructions
observed corresponding to a target function, then we choose
one out of $N$ dynamic instructions with a uniform random
probability of $\frac{1}{N}$ and flip the value of a randomly 
chosen bit of the destination virtual register, i.e.,
the l.h.s. of the randomly chosen dynamic instruction.
Similar error models have been proposed in the past for various 
resilience studies and it provides a reasonable estimate of 
application-level resiliency of an application\cite{llfi,kulfi13}.
Given that our focus is to study soft errors affecting 
\emph{structured address computation}, 
we consider all fault sites which when 
subjected to a random single-bit bit-flip may affect the output
of one or more \texttt{GEP} instructions of a target program.
Specifically, we propose two following error models which mainly 
differ in the dynamic fault site selection strategy.

\subsubsection{\textbf{Error Model I}}
As described in Sec.~\S\ref{sec:ex}, error scenario affecting 
\emph{structured addressed computations} are broadly categorized
into soft errors affecting \emph{index values} and the final output 
of \texttt{GEP} instructions.
Error model I considers the scenario where \emph{index values}
are corruption-free but the final output of one of the \texttt{GEP} 
instruction has a random single-bit corruption.
This is done by randomly choosing from dynamic instances of all GEP 
instructions of a target function, and injecting a bit-flip in the 
final address computed the \texttt{GEP} instruction.
\subsubsection{\textbf{Error Model II}}
Error model II considers the case where the \emph{index value}
of one of the dynamic instances of GEP instructions are corrupted
including the dynamic fault sites corresponding to the set of 
\emph{def-use} leading to the \emph{index-value} 

The above two error models are implemented using an open-source
and publicly available fault injector tool VULFI\cite{dpdns16vulfi,vulfiwebref}.
Also, note that in our error models, we do not target base addresses 
as these are small in numbers (one per array) and can be easily 
protected through replication without incurring severe performance 
or space overhead.

\subsection{\textbf{PRESAGE Transformations}}
We refer to two or more \texttt{GEP} instructions as 
\emph{same-class} \texttt{GEP}s if they use the
same base address.
PRESAGE creates a dependency chain between \emph{same-class}
\texttt{GEP}s in a two-stage process. 
\subsubsection{\textbf{Inter-Block Dependency Chains}}
The first stage involves enabling dependency chains between
\emph{same-class} \texttt{GEP}s in different basic blocks.
Intuitively, it would require first \texttt{GEP}, 
for a given base address, appearing in all basic blocks 
be transformed in a manner such that it uses the address 
computed by the last \emph{same-class} \texttt{GEP} in its 
predecessor basic block as the \emph{relative base}.
However, we need a bit more careful analysis as a basic block 
may have more than one predecessor basic blocks. 
Moreover, it might be possible that not all predecessor blocks
have a \emph{same-class} GEP or a predecessor block might be
a \emph{back edge} (i.e., there is a loop enclosing the basic block
and its predecessor basic block).
Therefore, we propose a three-step process for linking 
\emph{same-class} \texttt{GEP}s in different basic blocks 
as explained by Figs.~\ref{fig:igep1} and~\ref{fig:igep2}.
\begin{figure}[!h]
\begin{algorithmic}[1]
\Procedure{CreateInterBlkDepChain($\mathcal{F}$,$\mathcal{M}_G$,$\mathcal{M}_\phi$)}{}
\ForAll{$\mathcal{B}$ in BFS($\mathcal{F}$)}  
  \State $e$ $\gets$ GetIncomingEdgeCount($\mathcal{B}$)  
  \ForAll{$\mathbf{b}$ in $\mathcal{L}_\mathbf{b}(\mathcal{F})$}  
    \State $\phi$ $\gets$ CreateEmptyPHINode($\mathbf{b}$,$e$)
    \State InsertPHINodeEntry($\mathcal{B}$,$\mathbf{b}$,$\phi$,$\mathcal{M}_\phi$)
    \ForAll{$\mathcal{B}_p$ in $\mathcal{L}_{\mathcal{B}_p}(\mathcal{B})$}  
      \If{ HasGEP($\mathcal{B}_p$,$\mathbf{b}$,$\mathcal{M}_G$)}
	\State $\mathcal{G}$ $\gets$ GetGEP($\mathcal{B}_p$,$\mathbf{b}$,$\mathcal{M}_G$)
	\State SetIncomingEdge($\mathcal{B}_p$,$\mathcal{B}$,$\phi$,$\mathcal{G}$)
      \EndIf
    \EndFor  
  \EndFor
\EndFor
\EndProcedure
\end{algorithmic}
\caption{Creating Inter-Block Dependency Chains}
\label{fig:igep1}
\end{figure}

As a first step, as shown in Fig.~\ref{fig:igep1}, we iterate over 
all basic blocks of a target function $\mathcal{F}$ in a breadth-first 
order.
In a given basic block $\mathcal{B}$ with an incoming edge count
$e$, we insert a \texttt{phi} node for each unique base address
appearing in $\mathcal{L}_\mathbf{b}(\mathcal{F})$ for 
selecting a value from \emph{same-class} incoming \texttt{GEP} values 
(each belonging to a unique predecessor basic block).
For a given base address $\mathbf{b}$, the respective \texttt{phi} node 
entry is used as the \emph{relative base} by the first \texttt{GEP} 
(with base $\mathbf{b}$) in the current basic block $\mathcal{B}$.
In case, $\mathcal{B}$ does not have a valid \texttt{GEP} entry 
for $\mathbf{b}$ , then we call $\mathcal{B}$ a \emph{pass-through} basic 
block with respect to $\mathbf{b}$.
In this case, we simply pass the \texttt{phi} node value to the 
successor basic blocks.

We use a \texttt{phi} node because all PRESAGE transformations are applied 
at LLVM IR and LLVM uses the single static assignment (SSA) form thus
requiring a \texttt{phi} node to select a value from one or more 
incoming values.
For each \texttt{phi} node entry created in $\mathcal{B}$, if valid incoming 
\texttt{GEP} values are available from one or more predecessor basic blocks, 
the \texttt{phi} node is updated with those values by calling the 
\emph{SetIncomingEdge} routine.

\begin{figure}[!h]
\begin{algorithmic}[1]
\Procedure{UpdateInterBlkDepChain($\mathcal{F}$,$\mathcal{M}_\phi$,$\mathcal{M}_\mathcal{G}$,$P$)}{}
\ForAll{$\mathcal{B}$ in BFS($\mathcal{F}$)}  
  \ForAll{$\mathcal{B}_p$ in $\mathcal{L}_{\mathcal{B}_p}(\mathcal{B})$}  
    \ForAll{$\mathbf{b}$ in $\mathcal{L}_\mathbf{b}(\mathcal{F})$}        
      \State $s$ $\gets$ $\neg$HasGEP($\mathcal{B}_p$,$\mathbf{b}$,$\mathcal{M}_G$)
      \State $s$ $\gets$ $s$  $\wedge$ HasPHI($\mathcal{B}_p$,$\mathbf{b}$,$\mathcal{M}_\phi$)
      \State $s_1$ $\gets$ $s$ $\wedge$ $\neg$IsBackEdge($\mathcal{B}_p$,$\mathcal{B}$)
      \State $s_1$ $\gets$ $s_1$ $\wedge$ $(P=Pass1)$
      \State $s_2$ $\gets$ $s$ $\wedge$ IsBackEdge($\mathcal{B}_p$,$\mathcal{B}$)
      \State $s_2$ $\gets$ $s_2$ $\wedge$ $(P=Pass2)$
      \If{$s_1$ $\vee$ $s_2$}
	\State $\phi$ $\gets$ GetPHINode($\mathcal{B}$,$\mathbf{b}$,$\mathcal{M}_\phi$)      
	\State $\phi_p$ $\gets$ GetPHINode($\mathcal{B}_p$,$\mathbf{b}$,$\mathcal{M}_\phi$)
	\State SetIncomingEdge($\mathcal{B}_p$,$\mathcal{B}$,$\phi$,$\phi_p$)
      \EndIf
    \EndFor  
  \EndFor
\EndFor
\EndProcedure
\end{algorithmic}
\caption{Updating Inter-Block Dependency Chains}
\label{fig:igep2}
\end{figure}

At this point, we already have created \texttt{phi} node entries in each 
basic block (including all \emph{pass-through} basic blocks), and have populated 
these \texttt{phi} nodes with incoming \texttt{GEP} values wherever applicable.
As a next step, as shown in Fig.~\ref{fig:igep2}, for a basic block $\mathcal{B}$ 
with each of its \emph{pass-through} predecessor basic block with respect
to a base address $\mathbf{b}$, the respective \texttt{phi} node
$\phi$ is updated with the predecessor's \texttt{phi} node entry $\phi_p$
by calling \emph{SetIncomingEdge} routine.
If a back-edge exists from a \emph{pass-through} predecessor 
basic block $\mathcal{B}_p$ to $\mathcal{B}$ (i.e., there is exists a loop 
enclosing $\mathcal{B}$ and $\mathcal{B}_p$) then $\mathcal{B}$ may 
receive invalid data from $\mathcal{B}_p$ as $\mathcal{B}_p$ is also 
successor basic block of $\mathcal{B}$.
Therefore, we invoke the procedure \emph{UpdateInterBlkDepChain} in 
Fig.~\ref{fig:igep2} twice.
In the first pass, the \texttt{phi} node entries  of all \emph{pass-through} 
predecessor basic blocks of $\mathcal{B}$ which do not have back edges to 
$\mathcal{B}$, are assigned to the respective \texttt{phi} node entries in
$\mathcal{B}$.
In the second pass, we repeat the steps of the first pass with the exception
that this time we select the \texttt{phi} node entries  of all \emph{pass-through} 
predecessor basic blocks of $\mathcal{B}$ which do have back edges to 
$\mathcal{B}$.
\begin{figure}[!h]
\begin{algorithmic}[1]
\Procedure{CreateIntraBlkDepChain($\mathcal{F}$,$\mathcal{M}_G$,$\mathcal{M}_\phi$)}{}
\ForAll{$\mathcal{B}$ in BFS($\mathcal{F}$)}  
  \ForAll{$\mathbf{b}$ in $\mathcal{L}_\mathbf{b}(\mathcal{F})$}  
    \ForAll{$\mathcal{G}$ in $\mathcal{L}_\mathcal{G}(\mathcal{B},\mathbf{b})$}  
      \If{IsFirstGEP($\mathcal{G}$)}
	\State $\phi$ $\gets$ GetPHINode($\mathcal{B}$,$\mathbf{b}$,$\mathcal{M}_\phi$)
	\State $\mathbf{b}_r$ $\gets$ GetRelativeBase($\phi$,$\mathbf{b}$)
	\State $pid$ $\gets$ GetPrevIdx($\phi$)  
      \EndIf
      \State $id$ $\gets$ GetCurrentIdx($\mathcal{G}$)
      \State $rid$ $\gets$ GetRelativeIdx($id$,$pid$)
      \State $\mathcal{G}_n$ $\gets$ CreateNewGEP($\gamma$,$rid$)
      \State $pid$ $\gets$ $id$
      \State InsertGEP($\mathcal{G}_n$,$\mathcal{G}$)
      \State ReplaceAllUses($\mathcal{G}$,$\mathcal{G}_n$)    
      \State DeleteGEP($\mathcal{G}$)  
    \EndFor
  \EndFor
\EndFor
\EndProcedure
\end{algorithmic}
\caption{Creating Intra-Block Dependency Chains}
\label{fig:igep3}
\end{figure}
\subsubsection{\textbf{Intra-Block Dependency Chains}}
The second stage involves creating intra-block dependency chains.
As shown in Fig.~\ref{fig:igep3}, for each basic block $\mathcal{B}$ of 
a target function $\mathcal{F}$ and for each unique base address $\mathbf{b}$ 
$\in$ $\mathcal{L}_\mathbf{b}(\mathcal{B})$, if there exist one or more 
\emph{same-class} \texttt{GEP} instructions which use $\mathbf{b}$ as the base, 
we need to transform these \texttt{GEP}s to create a dependency chain.
In other words, each \texttt{GEP} uses the value computed by the previous 
\texttt{GEP} as the \emph{relative base} using our RBA scheme as explained 
in Eq.~\ref{eq:rba}.
For the first occurrence of \texttt{GEP} instruction in $\mathcal{B}$ 
with base $\mathbf{b}$, we extract the \emph{relative base} information
using the \emph{phi} node entry $\phi$ created in the previous stage.
At runtime, the \texttt{phi} node $\phi$ will receive the last address
computed using the base address $\mathbf{b}$ from one of the predecessor
basic blocks of $\mathcal{B}$.
In summary, for each \texttt{GEP} instruction $\mathcal{G}$, an equivalent 
version  $\mathcal{G}_n$ is created using the \emph{relative base} and
the \emph{relative index} values. 
All uses of $\mathcal{G}$ are then replaced by $\mathcal{G}_n$ and
$\mathcal{G}$ is then finally deleted.
\subsection{\textbf{Detector Design}}
\begin{figure}[!h]
\begin{algorithmic}[1]
\Procedure{InsertDetectors($\mathcal{F}$,$\mathcal{M}_G$,$\mathcal{M}_\phi$)}{}
\ForAll{$\mathcal{B}_e$ in $\mathcal{L}_{\mathcal{B}_e}(\mathcal{F})$}
  \ForAll{$\mathbf{b}$ in $\mathcal{L}_\mathbf{b}$}
    \State $\phi$ $\gets$ GetPHINode($\mathcal{B}$,$\mathbf{b}$,$\mathcal{M}_\phi$)
    \State $\mathbf{b}_r$ $\gets$ GetRelativeBase($\phi$,$\mathbf{b}$)
    \State $rid$ $\gets$ GetRelativeIdx($\phi$)  
    \State $pid$ $\gets$ GetPrevIdx($\phi$)  
    \State $\mathcal{G}$ $\gets$ CreateNewGEP($\mathbf{b}_r$,$rid$)
    \State $\mathcal{G}_d$ $\gets$ CreateNewGEP($\mathbf{b}$,$pid$)
    \State InsertEqvCheck($\mathcal{G}$,$\mathcal{G}_d$)
  \EndFor
\EndFor
\EndProcedure
\end{algorithmic}
\caption{Algorithm for Error Detectors}
\label{alg:det1}
\end{figure}
The error detectors are designed to protect against single-bit faults injected
using error model I.
As shown in Fig.~\ref{alg:det1}, in each exit basic block $\mathcal{B}_e$,
for each unique base address $\mathbf{b}$, PRESAGE makes available the value 
computed of the last run \texttt{GEP} instruction with base $\mathbf{b}$ and the 
\emph{relative index} value used.
Additionally, PRESAGE also makes available the \emph{absolute index} value which 
along with the base address $\mathbf{b}$ can also be used to reproduce the output
of the last run \texttt{GEP} instruction with base $\mathbf{b}$.
The error detectors then simply check if the output $\mathcal{G}$ produced by
the last run \texttt{GEP} instruction matches the recomputed value $\mathcal{G}_d$ 
using the base address $\mathbf{b}$ and the \emph{absolute index} value.
Given that in error model I, we consider the base address and index value to 
be corruption free, the error detectors are \emph{precise} with respect to error 
model I as they do not report any \emph{false positives}.
\begin{figure*}
\begin{minipage}{0.5\textwidth}
\includegraphics[width=6cm,height=7.5cm]{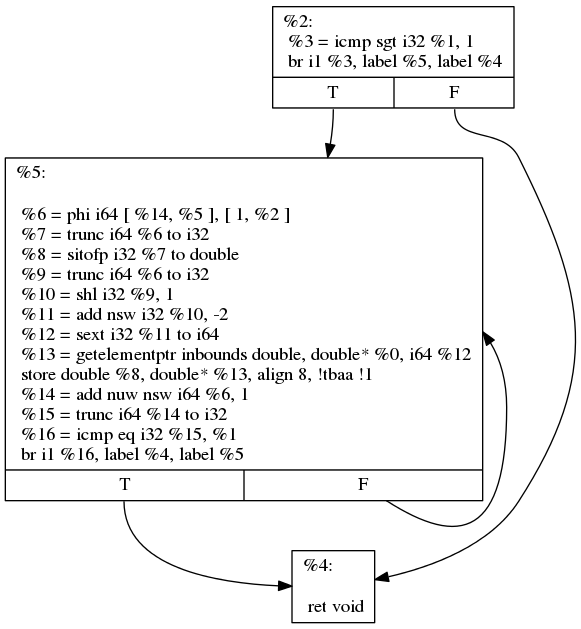}
\caption{LLVM IR level CFG representation of the function \texttt{foo1}}
\label{cfg:foo1}
\end{minipage}
\begin{minipage}{0.5\textwidth}
\includegraphics[width=6.5cm,height=8.5cm]{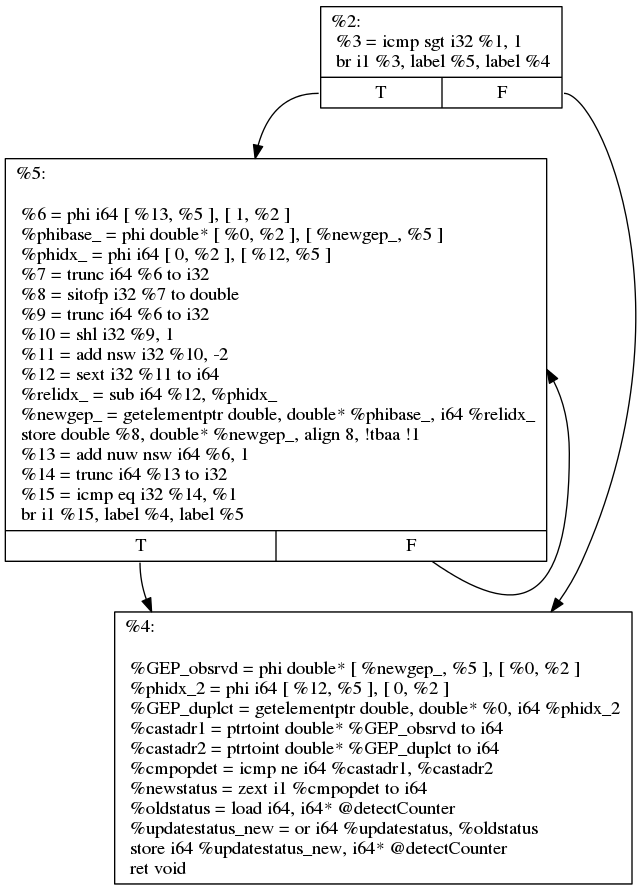}
\caption{LLVM IR level CFG representation of PRESAGE transformed version of the function \texttt{foo1}}
\label{cfg:foo1_psg}
\end{minipage}
\end{figure*}

Fig.~\ref{cfg:foo1} shows the LLVM-level control-flow graph (CFG) of the function \texttt{foo1}
presented in Sec.\S\ref{sec:ex}.
Similarly, Fig.~\ref{cfg:foo1_psg} shows the LLVM-level CFG of the PRESAGE 
transformed version of the function \texttt{foo1}.
The \texttt{GEP} instruction in function \texttt{foo1} (Fig.~\ref{cfg:foo1}) which stores the computed address
in register \texttt{\%13} is replaced by a new \texttt{GEP} instruction (Fig.~\ref{cfg:foo1_psg}) in 
the PRESAGE transformed version of \texttt{foo1} which uses relative base and relative index value for 
address computation.
The PRESAGE transformed version of \texttt{foo1} in Fig.~\ref{cfg:foo1} also has error detector
code inserted in the exit basic block.
Specifically, \texttt{\%GEP\_duplct} represents the recomputed version of the address which is 
compared against the observed address value \texttt{\%GEP\_obsrvd}. 
In case of a mismatch, the global variable \texttt{@detectCounter} is set to report error 
detection to the end user.

\section{\textbf{Experimental Results}}
\label{sec:result}
\label{sec:result}
\subsection{\textbf{Evaluation Strategy}}
Our evaluation strategy involves measuring the effectiveness
of the proposed error detectors in terms of
SDC detection rate and performance overhead.
In addition, we analyze the impact of 
PRESAGE transformations on an application's resiliency using a
fault injection driven study.
We consider 10 benchmarks listed in 
Table\ref{tab:bench} drawn from the PolyBench/C benchmark 
suite\cite{polybench}.
These benchmarks represent a diverse set of applications from 
areas such as stencils, algebraic kernels, solvers, and BLAS 
routines.
\begin{table}[!h]
 \centering
 \begin{tabular}{|c||l|}
 \hline
 \rule{0pt}{2ex} \textbf{Experiment Set} & \textbf{Description} \\
 \hline
 Native\_FIC\_EM-I & \rule{0pt}{6ex} \shortstack[l]{A fault-injection campaign (FIC) \\
  using error model I on the native\\ version of a target benchmark.}\\
 \hline
 Native\_FIC\_EM-II & \rule{0pt}{5ex} \shortstack[l]{Same as Native\_FIC\_EM-I except \\
 that error model II is used.}\\
 \hline
 Presage\_FI\_EM-I & \rule{0pt}{5ex} \shortstack[l]{A fault-injection campaign (FIC) \\
 using error model I on benchmarks \\ transformed using PRESAGE.}\\
 \hline
 Presage\_FI\_EM-II & \rule{0pt}{5ex} \shortstack[l]{Same as Presage\_FI\_EM-I except \\
 that error model II is used.}\\
 \hline
 \end{tabular}
 \caption{Summary of experiments}
 \label{tab:expset}
 \vspace{-1em}
\end{table}
For each of these benchmarks, we perform four set of experiments, 
summarized in Table\ref{tab:expset}.
Each experiment set involves a fault injection campaign (FIC),
consisting of 5000 independent \emph{experimental runs}.
In each \emph{experimental run}, we carry out a \emph{fault-free} and 
a \emph{faulty} execution of a target benchmark using identical
program input parameters and compare the outcome of the
two executions.
The program input parameters (such as array size used in the 
benchmark) are randomly chosen from a predefined range of values.
During a \emph{fault-free} execution, no faults are injected whereas 
during a \emph{faulty} execution, a single-bit fault is injected
in a dynamic LLVM IR instruction selected randomly 
using either error model I or error model II as explained in 
Sec.~\S\ref{sec:method}.

Note that we only target the key function(s) that implement the core logic 
of a benchmark for fault injection.
For example, in the \texttt{jacobi-2d} benchmark, we only target the 
\emph{kernel\_jacobi\_2d} which implements the core jacobi kernel and
ignore the other auxiliary functions such as the function used for array 
initialization or the program's \emph{main()}.

Given that the benchmarks chosen produce one or more \emph{result arrays} as the final 
program output, we compare respective elements of the \emph{result arrays} produced 
by the \emph{faulty} and \emph{fault-free} executions to categorize the outcome
of the \emph{experimental run} 
as:

\noindent
\textbf{SDC}: The executions ran to completion, but the corresponding elements of the \emph{result arrays} of the
\emph{fault-free} and \emph{faulty} execution are not 
equivalent.

\noindent
\textbf{Benign}: The correponding elements of the \emph{result arrays} of the
\emph{fault-free} and \emph{faulty} execution are equivalent.

\noindent
\textbf{Program Crash}: The program crashes or terminates prematurely 
without producing the final output.

We analyze the impact of PRESAGE transformations on an application's
resiliency by comparing the outcomes of the experiment sets
Native\_FIC\_EM I with Presage\_FI\_EM-I, and Native\_FIC\_EM-II 
with Presage\_FI\_EM-II.
\begin{figure*}[!t]
\includegraphics[width=18cm,height=5cm]{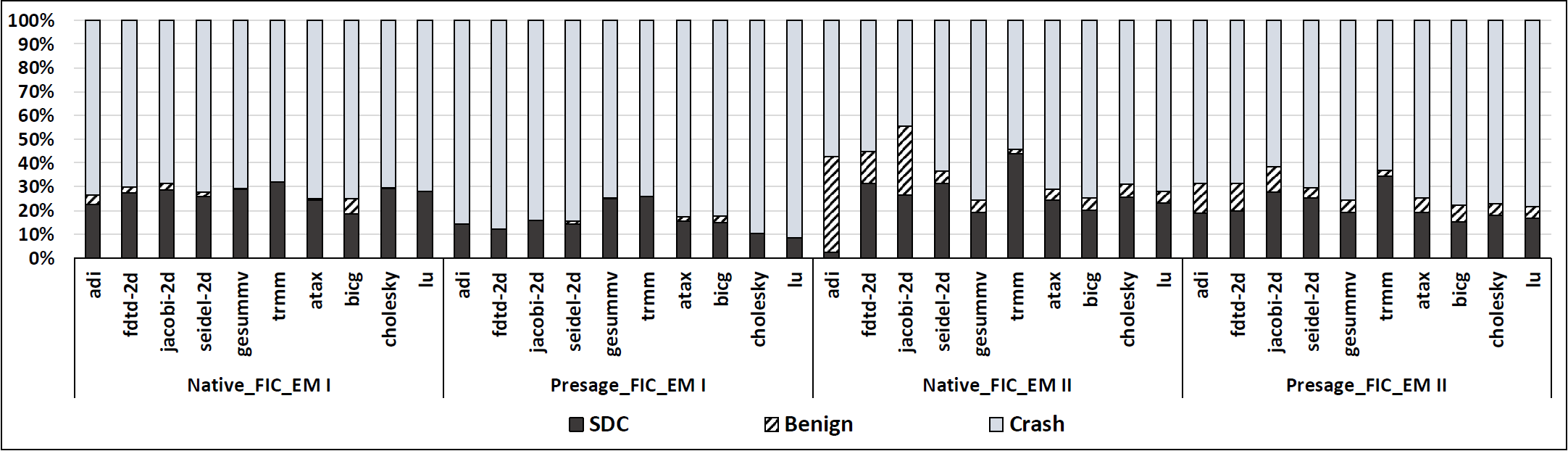}
\caption{Outcomes of the fault injection campaigns}
\label{fig:fic}
\end{figure*}
\subsection{\textbf{Fault Injection Campaigns}}
Fig.\ref{fig:fic} shows the result of FIC done under each
experiment set listed in Table\ref{tab:expset}.
Each column in the figure represents an FIC consisting of 
5000 runs.
Therefore, the total number of fault injections done across 10 
benchmarks and 4 experiment sets stands at 0.2 million (4 experiment 
sets $\times$ 10 benchmarks $\times$ 5000 fault injections).

\noindent
\textbf{Non-trivial SDC Rates}: The results for experiment sets
Native\_FIC\_EM-I and Native\_FIC\_EM-II shown in Fig.\ref{fig:fic} 
demonstrate that non-trivial SDC rates are observed when 
\emph{structured address computations} are subjected to bit flips.
Specifically, for the experiment set Native\_FIC\_EM-I, we observe a 
maximum and a minimum SDC rates of 32.2\% and 18.5\%,
for the benchmarks \texttt{trmm} and \texttt{bicg}, respectively.
In case of Native\_FIC\_EM-II, we observe a greater contrast,
with a maximum SDC rate of 43.6\% and a minimum SDC rate of 2.3\% 
for the benchmarks \texttt{trmm} and \texttt{adi}, respectively.

\noindent
\textbf{Promotion of SDCs to Program Crashes}: When comparing the results
of experiment sets Presage\_FI\_EM-I and Presage\_FI\_EM-II with
that of Native\_FIC\_EM-I and Native\_FIC\_EM-II, we observe
that PRESAGE transformations lead to a sizable fraction of SDCs 
getting promoted to program crashes.
Specifically, Presage\_FI\_EM-I reports an average increase of
12.5\% (averaged across all 10 benchmarks) in the number of program 
crashes when compared to Native\_FI\_EM-I, with a maximum increase
of 19.3\% reported for the \texttt{cholesky} benchmark.
Similarly, Presage\_FI\_EM-II reports an average increase of
7.8\% (averaged across all 10 benchmarks) in the number of program 
crashes when compared to Native\_FI\_EM-II with a maximum increase
of 16.8\% reported for the \texttt{jacobi-2d} benchmark.

\subsection{\textbf{Detection Rate \& Performance Overhead}}
\begin{figure}[!h]
\includegraphics[width=8.5cm,height=5.5cm]{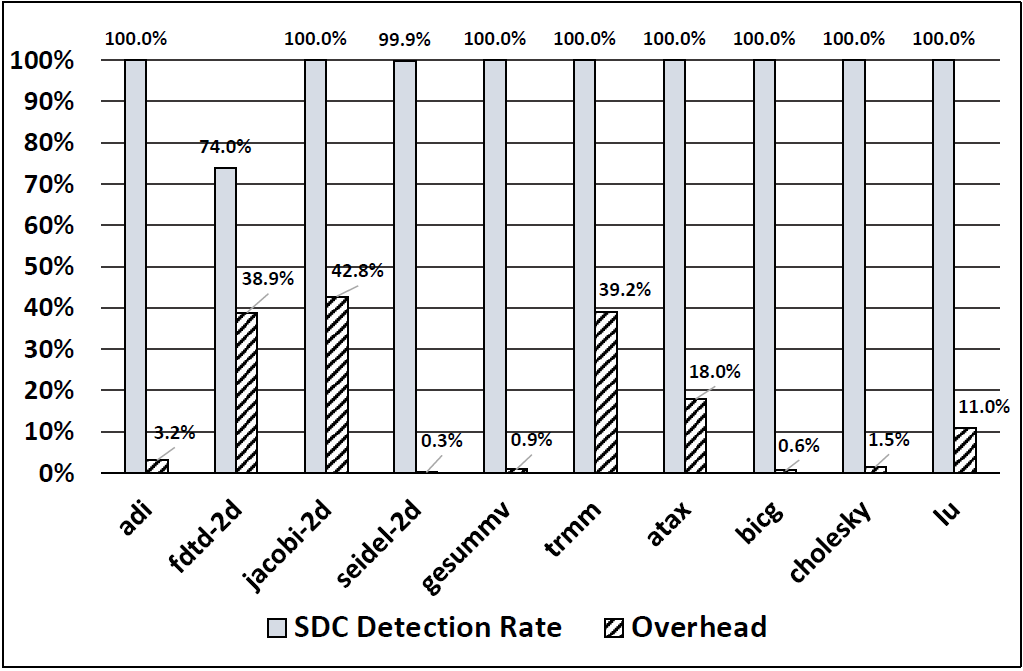}
\caption{SDC detection rate \& performance overhead}
\label{fig:dtr_ovr}
\end{figure}
Fig.\ref{fig:dtr_ovr} shows the percentage of SDCs
reported in Fig.\ref{fig:fic} under 
Presage\_FIC\_EM-I that are detected by the PRESAGE-inserted 
error detectors. 
Except for the benchmark \texttt{fdtd-2d}, we are able to 
detect 100\% of the SDCs caused by a random bit-flip
injected using error model I.
In case of \texttt{fdtd-2d}, we are able to detect only 
74\% of the reported SDCs because a fraction of \texttt{GEP}
instructions in \texttt{fdtd-2d} have mutable base addresses.
Recall that the PRESAGE transformations can only be applied
to \texttt{GEP} instructions with immutable base 
addresses.

For the benchmarks \texttt{adi}, \texttt{seidel-2d}, \texttt{gesummv}, 
\texttt{bicg}, and \texttt{cholesky}, we notice that the error detectors
incur almost negligible overheads ranging between 0.3\% and 3.2\%.
Benchmarks \texttt{lu} and \texttt{atax} report overhead figures
of less than 20\% whereas the benchmarks \texttt{jacobi-2d}, 
\texttt{fdtd-2d}, \texttt{trmm}, and \texttt{atax} report 
overhead figures of close to 40\%.

Upon further investigation (details omitted for brevity),
we found that the higher overheads for some of the 
benchmarks are due to the dependency chains (especially 
inter-block dependency chains) introduced by the PRESAGE 
transformations.
Recall that due to these dependency chains, 
a \emph{structured address computation} in a basic block 
might require a previously computed address from one of its predecessor 
basic blocks.
Due to this dependency, a register value may have to be kept 
alive for a longer duration, increasing the chances of 
a register spill, which in turn leads to a higher overhead.
Therefore, we recommend additional care while deploying 
PRESAGE on any new program by first checking (on small test inputs) 
if the 
introduced overhead is within acceptable limits for 
the end user.
Finding an optimal set of dependency chains or splitting a dependency
chain into smaller ones to provide the best trade-off between the
detection rate and overhead is beyond the scope of this paper and
constitutes future work.

\subsection{\textbf{False Positives \& False Negatives}}
We refer to the errors flagged during the execution of a 
PRESAGE-transformed program as a false positive when no faults
are injected during the execution.
Conversely, if there are no errors reported during the execution 
of a PRESAGE-transformed program while an error is actually injected 
during the execution, then we regard it as a false negative.

The basic philosophy of the PRESAGE detectors is to 
recompute the final address observed at the end point of a 
dependency chain and compare the recomputed address against 
the observed final address.
Also, in error model I, the index and the base address 
of a \texttt{GEP} instruction are assumed to be corruption-free 
but the final address computed by it can be erroneous.
Therefore, under error model I, whenever the recomputed address 
does not match the observed address, it 
attributes it to an actual bit-flip.
In summary, the detectors never report false positives
under error model I.
Even in the case of error model II, where we subject the index
value of a \texttt{GEP} instruction to a bit-flip, the value recomputed 
by the detectors would use the same corrupted index value to reproduce 
the same corrupted observed value.
Thus even under error model II, the error detectors must not 
report false positive. However, it may report false negatives,
including in cases 
where we inject bif flips into \texttt{GEP} instructions 
that have mutable base addresses, as in the case of 
\texttt{fdtd-2d} benchmark.

\subsection{\textbf{Coverage Analysis}}
Table~\ref{tab:bench} provides an insight into the kind 
of coverage provided by the PRESAGE-based error detectors.
Total SIC denotes the total static instruction count of
the LLVM IR instructions corresponding to key function(s)
of a benchmark that are targeted for fault injections.
SIC-I and SIC-II represents the subset of instructions 
represented by SIC chosen using error model I and 
error model II respectively.
Clearly, SIC-I and SIC-II represent a significant portion
of SIC with the share of SIC-I ranging between 15.5\% and 
28.5\% where as that of SIC-II ranging between 63.3\% and 
21.7\%.
The ratio between SIC-I and SIC-II roughly varies from 
1:3 (in case of \texttt{seidel-2d}) to 1:1 (in case of 
\texttt{gesummv} and \texttt{bicg}).
Avg. DIC-I is a counterpart of SIC-I, representing
the average dynamic instruction count averaged over DIC 
observed during each experimental run of an FIC done under 
the experiment set Native\_FIC\_EM-I.
Similarly, Avg. DIC-II denotes the average dynamic instruction 
count averaged over DIC observed during each experimental run 
of an FIC done under the experiment set Native\_FIC\_EM-II.
Clearly, the fault sites considered under error model I and II
constitute a significant part of the overall static instruction count of 
the benchmarks considered in our experiments.

\begin{table*}[!t]
 \centering
 \begin{tabular}{|c||c|c|c|c|c|c|c|}
 \hline
 \rule{0pt}{4ex} \textbf{Benchmark} & 
 \rule{0pt}{5ex} \shortstack{\textbf{Avg. DIC-I}\\(in millions)} & \shortstack{\textbf{Avg. DIC-II}\\(in millions)} &  
 \rule{0pt}{5ex} \textbf{SIC-I} & \textbf{SIC-II} & \textbf{\shortstack{Total\\SIC}} & \textbf{\%SI-I} & \textbf{\%SI-II} \\
 \hline    
 adi & 59.2 & 157.5 & 30 & 69 & 161 & 18.6\% & 42.8\%  \\
 \hline
 fdtd-2d & 63.7 & 24.8 & 68 & 98 & 249 & 27.3\% & 39.3\% \\
 \hline
 seidel-2d & 74.8 & 36.8  & 42 & 114 & 180 & 23.3\% & 63.3\% \\
 \hline
 jacobi-2d & 64.2 & 97.1  & 56 & 112 & 196 & 28.5\% & 57.1\% \\
 \hline
 gesummv & 0.4 & 0.7 &  5 & 5 & 22 & 22.7\% & 22.7\%  \\
 \hline
 trmm & 39.1 & 107.1 &  14 & 39 & 90 & 15.5\% & 43.3\% \\
 \hline 
 atax & 0.5 & 0.7 &  22 & 26 & 91 & 24.1\% &  28.5\% \\
 \hline 
 bicg & 0.4 & 0.7 &  5 & 5 & 23  & 21.7\% & 21.7\%  \\
 \hline 
 cholesky & 0.3 & 0.8 &  16 & 39 & 89 & 17.9\% & 43.8\% \\
 \hline
 lu & 0.6 & 1.9 &  15 & 35 & 77 & 19.4\% & 45.4\% \\
 \hline 
 \end{tabular}\rule{0pt}{4ex}
 \caption{Benchmark description}
 \label{tab:bench}
 \vspace{-1em}
\end{table*}

\section{\textbf{Conclusions \& Future work}}
\label{sec:conclusion}
\label{sec:con}
Researchers in the HPC community have highlighted the growing need for 
developing cross-layer resilience solutions with application-level 
techniques gaining a prominent place due to their inherent flexibility.  
Developing efficient light-weight error detectors has been a central
theme of application-level resilience research dealing with silent
data corruption.
Through this work, we argue that, often, protecting 
\emph{structured address computations} is important 
due to their vulnerability to bit flips, resulting in non-trivial SDC
rates.
We experimentally support this argument by 
carrying out fault injection driven experiments on 10 well-known 
benchmarks.
We witness SDC rates ranging between 18.5\% and 43.6\% when 
instructions in these benchmarks pertaining to \emph{structured 
address computations} are subjected to bit flips.

Next, guided by the principle that maximizing the propagation 
of errors would make them easier to detect, we introduce a novel
approach for rewriting the address computation logic used in
\emph{structured address computations}.
The rewriting scheme, dubbed the RBA scheme, introduces a dependency
chain in the address computation logic, enabling sufficient
propagation of any error and, thus, allowing efficient placement of
error detectors.
Another salient feature of this scheme is that it promotes 
a fraction of SDCs (user-invisible) to program crashes 
(user-visible).
One can argue that promoting SDCs to program crashes may lead
to a bad user experience. However, a program crash 
is far better than an SDC whose insidious nature does not
raise any user alarms while silently invalidating
the program output.

We have implemented our scheme as a compiler-level technique 
called PRESAGE developed using the LLVM compiler infrastructure.
In Sec.\S\ref{sec:method}, we formally presented the key steps 
involved in implementing the PRESAGE transformations which include
creating inter-block and intra-block dependency chains, and 
a light-weight detector placed strategically at all exit 
points of a program.
We reported high detection rates ranging between 74\% and 100\%
with the performance overhead ranging between 0.3\% and 42.8\%
across 10 benchmarks.
In addition, the PRESAGE-transformed benchmarks witness an average and a maximum 
increase of 12.5\% and 19.3\% respectively in program crashes as 
compared to their original versions when faults are injected using 
error model I.
These figures stands at 7.8\% and 16.8\% respectively when 
error model II is used instead of error model I for 
fault injections.

Our current work also identifies some challenges  
we plan to address as part of the future work.
Specifically, we observe a relatively higher detection overhead
for some of the benchmarks, due to increased
register pressure caused by the introduction of dependency 
chains as explained earlier.
In the  future, we plan to explore efficient ways of mining
\texttt{GEP} instructions in a program that are best 
suited for PRESAGE transformations without adversely impacting 
performance. 
Although, the main focus of our work is to provide coverage 
explicitly for error model I, we also observe that PRESAGE provides 
partial coverage for error model II by promoting a fraction of SDCs 
to program crashes.
As future work, we plan to explore techniques used in the context of 
verification and polyhedral transformations to
develop comprehensive error detection mechanisms for error model II.
Finally, through this work, we hope to bring to the resilience
community's notice the importance and the need for developing efficient 
error detectors for protecting \emph{structured address computations}.

\section{\textbf{Acknowledgement}}
\label{sec:ack}
This work was supported in part by the U.S. Department of Energy’s
(DOE) Office of Science, Office of Advanced Scientific Computing
Research, under award 66905. Pacific Northwest National Laboratory is
operated by Battelle for DOE under Contract DE-AC05-76RL01830.
The Utah authors were supported in part under the same DOE project with
award number 55800790, NSF Award CCF 1255776, and SRC Tasks 2425.001, 2426.001.

\bibliographystyle{IEEEtran}
\bibliography{IEEEabrv,UtahResilience}

\end{document}